\documentclass[prb,twocolumn,showpacs,preprintnumbers,superscriptaddress,amsmath,amssymb]{revtex4}

\usepackage[dvips]{color}

\usepackage{graphicx}
\usepackage{dcolumn}
\usepackage{bm}
\usepackage{latexsym}

\begin{document}

\title{Strong carrier-scattering in iron-pnictide superconductors with highest $T_{\rm c}$ obtained from charge transport experiments}

\author{S. Ishida}
 \affiliation{Department of Physics, University of Tokyo, Tokyo 113-0033, Japan}
 \affiliation{National Institute of Advanced Industrial Science and Technology, Tsukuba 305-8568, Japan}
 \affiliation{JST, Transformative Research-Project on Iron Pnictides, Tokyo 102-0075, Japan}
 
\author{M. Nakajima}
 \affiliation{Department of Physics, University of Tokyo, Tokyo 113-0033, Japan}
 \affiliation{National Institute of Advanced Industrial Science and Technology, Tsukuba 305-8568, Japan}
 \affiliation{JST, Transformative Research-Project on Iron Pnictides, Tokyo 102-0075, Japan}
 
\author{Y. Tomioka}
 \affiliation{National Institute of Advanced Industrial Science and Technology, Tsukuba 305-8568, Japan}
 \affiliation{JST, Transformative Research-Project on Iron Pnictides, Tokyo 102-0075, Japan}
 
\author{T. Ito}
 \affiliation{National Institute of Advanced Industrial Science and Technology, Tsukuba 305-8568, Japan}
 \affiliation{JST, Transformative Research-Project on Iron Pnictides, Tokyo 102-0075, Japan}
 
\author{K. Miyazawa}
 \affiliation{National Institute of Advanced Industrial Science and Technology, Tsukuba 305-8568, Japan}
 \affiliation{JST, Transformative Research-Project on Iron Pnictides, Tokyo 102-0075, Japan}
 \affiliation{Department of Applied Electronics, Tokyo University of Science, Chiba 278-3510, Japan}
 
\author{H. Kito}
 \affiliation{National Institute of Advanced Industrial Science and Technology, Tsukuba 305-8568, Japan}
 \affiliation{JST, Transformative Research-Project on Iron Pnictides, Tokyo 102-0075, Japan}
 
\author{C. H. Lee}
 \affiliation{National Institute of Advanced Industrial Science and Technology, Tsukuba 305-8568, Japan}
 \affiliation{JST, Transformative Research-Project on Iron Pnictides, Tokyo 102-0075, Japan}
 
\author{M. Ishikado}
 \affiliation{National Institute of Advanced Industrial Science and Technology, Tsukuba 305-8568, Japan}
 \affiliation{JST, Transformative Research-Project on Iron Pnictides, Tokyo 102-0075, Japan}
 \affiliation{Japan Atomic Energy Agency, Ibaraki, 319-1195, Japan}
 
\author{S. Shamoto}
 \affiliation{JST, Transformative Research-Project on Iron Pnictides, Tokyo 102-0075, Japan}
 \affiliation{Japan Atomic Energy Agency, Ibaraki, 319-1195, Japan}

\author{A. Iyo}
 \affiliation{National Institute of Advanced Industrial Science and Technology, Tsukuba 305-8568, Japan}
 \affiliation{JST, Transformative Research-Project on Iron Pnictides, Tokyo 102-0075, Japan}
 
\author{H. Eisaki}
 \affiliation{National Institute of Advanced Industrial Science and Technology, Tsukuba 305-8568, Japan}
 \affiliation{JST, Transformative Research-Project on Iron Pnictides, Tokyo 102-0075, Japan}
 
\author{K. M. Kojima}
 \affiliation{Department of Physics, University of Tokyo, Tokyo 113-0033, Japan}
 \affiliation{JST, Transformative Research-Project on Iron Pnictides, Tokyo 102-0075, Japan}

\author{S. Uchida}
 \affiliation{Department of Physics, University of Tokyo, Tokyo 113-0033, Japan}
 \affiliation{JST, Transformative Research-Project on Iron Pnictides, Tokyo 102-0075, Japan}

{\color{red}}

\begin{abstract}
\indent Characteristic normal-state charge transport is found in the oxygen-deficient iron-arsenides $Ln$FeAsO$_{1-y}$ ($Ln$: La and Nd) with the highest $T_{\rm c}$'s among known Fe-based superconductors. The effect of ``doping'' in this system is mainly on the carrier scattering, quite distinct from that in high-$T_{\rm c}$ cuprates. In the superconducting regime of the La system with maximum $T_{\rm c}$ = 28~K, the low-temperature resistivity is dominated by a $T^2$ term. On the other hand, in the Nd system with $T_{\rm c}$ higher than 40~K, the carriers are subject to stronger scattering showing $T$-linear resistivity and small magnetoresistance. Such strong scattering appears crucial for high-$T_{\rm c}$ superconductivity in the iron-based system.
\end{abstract}

\maketitle

\section{\label{sec:level1}Introduction}
\indent Among various classes of Fe-based superconductors, the highest superconducting transition temperature $T_{\rm c}$ exceeding 50~K is realized exclusively in the $Ln$FeAsO(F) system ($Ln$: lanthanide elements).~\cite{Kamihara,X.H.Chen,Rotter,Pitcher,Hsu} It is empirically known that $T_{\rm c}$ of $Ln$FeAsO correlates strongly with the crystal structure and becomes maximum when FeAs$_4$ forms a nearly ideal tetrahedron, namely, when the bond angle of As-Fe-As, $\alpha$, is $\sim$ 109.5$^{\mathrm{o}}$.~\cite{C.H.Lee} However, it is still an unresolved puzzle why $T_{\rm c}$ is so different among different classes. Even in $Ln$FeAsO(F), the optimal $T_{\rm c}$ is 28~K for $Ln$ = La, whereas for $Ln$ = Nd, Sm, Gd, Tb, and Dy, the maximum $T_{\rm c}$ is higher than 50~K.~\cite{Miyazawa} To understand this difference in $T_{\rm c}$, it is desirable to search for its origin in the electronic structure, as well as in the electron dynamics.

\indent Recall that in the conventional BCS superconductors, $T_{\rm c}$ is more or less connected to scattering of the carriers in the normal-state charge transport, since the electron--phonon interaction is the origin of both the electron pairing that determines $T_{\rm c}$ and the carrier scattering that determines the resistivity.~\cite{Gurvitch} For the high-$T_{\rm c}$ cuprate superconductors, the carrier scattering rate is basically proportional to temperature ($T$), and the $T$-linear scattering rate is nearly independent of material and doping. This ``universal'' charge transport is considered to be a hallmark of the high-$T_{\rm c}$ cuprates. As $T_{\rm c}$ is strongly material dependent, it is not directly correlated with the scattering.

\indent Here, we investigate the transport properties of the oxygen-deficient oxypnictides LaFeAsO$_{1-y}$ and NdFeAsO$_{1-y}$ to explore how the charge transport in the normal state is correlated with the superconducting $T_{\rm c}$ in this class of pnictides, as well as in other classes of FeAs-based superconductors for comparison. Recently, single crystals of doped BaFe$_2$As$_2$ systems have been synthesized, and their transport properties have been investigated.~\cite{L.Fang,Rullier-Albenque,Kasahara} Since single crystals of the $Ln$FeAsO system are not available for a systematic study, we measured polycrystalline samples. To the best of our knowledge, this is the first systematic study of oxygen-deficient $Ln$FeAsO$_{1-y}$, covering a wide range of doping levels and revealing the evolution of the transport coefficients, resistivity, Hall coefficient, and magnetoresistance. Since the oxypnictides are known as a multiband system with a Fermi surface consisting of three hole and two electron sheets, we have to take this into account in analyzing the transport properties by measuring the temperature and magnetic-field dependences of these transport coefficients. In this paper, we will show that $T_{\rm c}$ seems to be correlated with the scattering in the case of Fe-based superconductors.


\section{\label{sec:level2}Experimental methods}
\indent Polycrystalline samples of $Ln$FeAsO$_{1-y}$ were prepared by a high-pressure synthesis technique using a cubic-anvil-type apparatus (Riken CAP-07).~\cite{Kito} Well-sintered samples of high density, sufficient for transport measurements, were obtained by applying a pressure of 2~GPa during the synthesis. We found no difference in the sample quality between $Ln$ = La and Nd. Because some of the starting materials are quite reactive to oxygen, we could not prevent their oxidation in the synthesis procedure. The actual amount of oxygen deficiency $y$ was estimated with reference to the lattice constants~\cite{C.H.Lee} which were slightly smaller than those of the nominal composition. There was no trace of impurity phases in x-ray diffraction patterns. In order to make the distribution of the oxygen deficiency homogeneous, the samples were annealed in air at 450~$^{\mathrm{o}}$C. The introduction of oxygen deficiency is considered to work as electron doping, which seems equivalent to the substitution of fluorine for oxygen.~\cite{C.H.Lee,Miyazawa} A standard four-terminal method was used for the resistivity measurements. Since the anisotropy ratio $\rho_c$/$\rho_{ab}$ is estimated to be $\sim$25 or larger,~\cite{Jia,Ishikado} it is reasonable to assume that the transport properties of the polycrystalline samples mainly reflect the in-plane ones. From comparison of the transport data of polycrystalline samples and those of single crystals of other Fe-pnictide systems, such as Ba(Fe$_{1-x}$Co$_x$)$_2$As$_2$,~\cite{L.Fang,Rullier-Albenque}, and also from the result for single crystalline PrFeAsO$_{1-y}$,~\cite{Ishikado}, in these dense polycrystalline samples obtained by high-pressure synthesis, the grain boundary contributions to both resistivity and Hall coefficient are small, scarcely affecting their $T$- and $H$-dependences. The Hall resistivity and the transverse MR were measured in a cryostat equipped with a superconducting magnet (PPMS, Quantum Design Inc.) with the magnetic field normal to the widest plane.

\section{\label{sec:level3}Results and discussions}
\indent For the La system, superconductivity appears around $y \sim$ 0.08 (``underdoped'' regime), $T_{\rm c}$ becomes maximum (28~K) around $y \sim$ 0.11 (``optimally'' doped), and $T_{\rm c}$ decreases slowly for $y$ $>$ 0.11 in the ``overdoped'' regime. We name each doping regime analogously to the convention used for high-$T_{\rm c}$ cuprate superconductors for convenience. For the Nd system, $T_{\rm c}$ jumps up to above 40~K at $y \sim$ 0.11, and then $T_{\rm c}$ increased gradually and reaches 52~K at $y$ = 0.17, the highest $y$ we prepared.

\begin{figure}[t!]
\includegraphics[width=\columnwidth,clip]{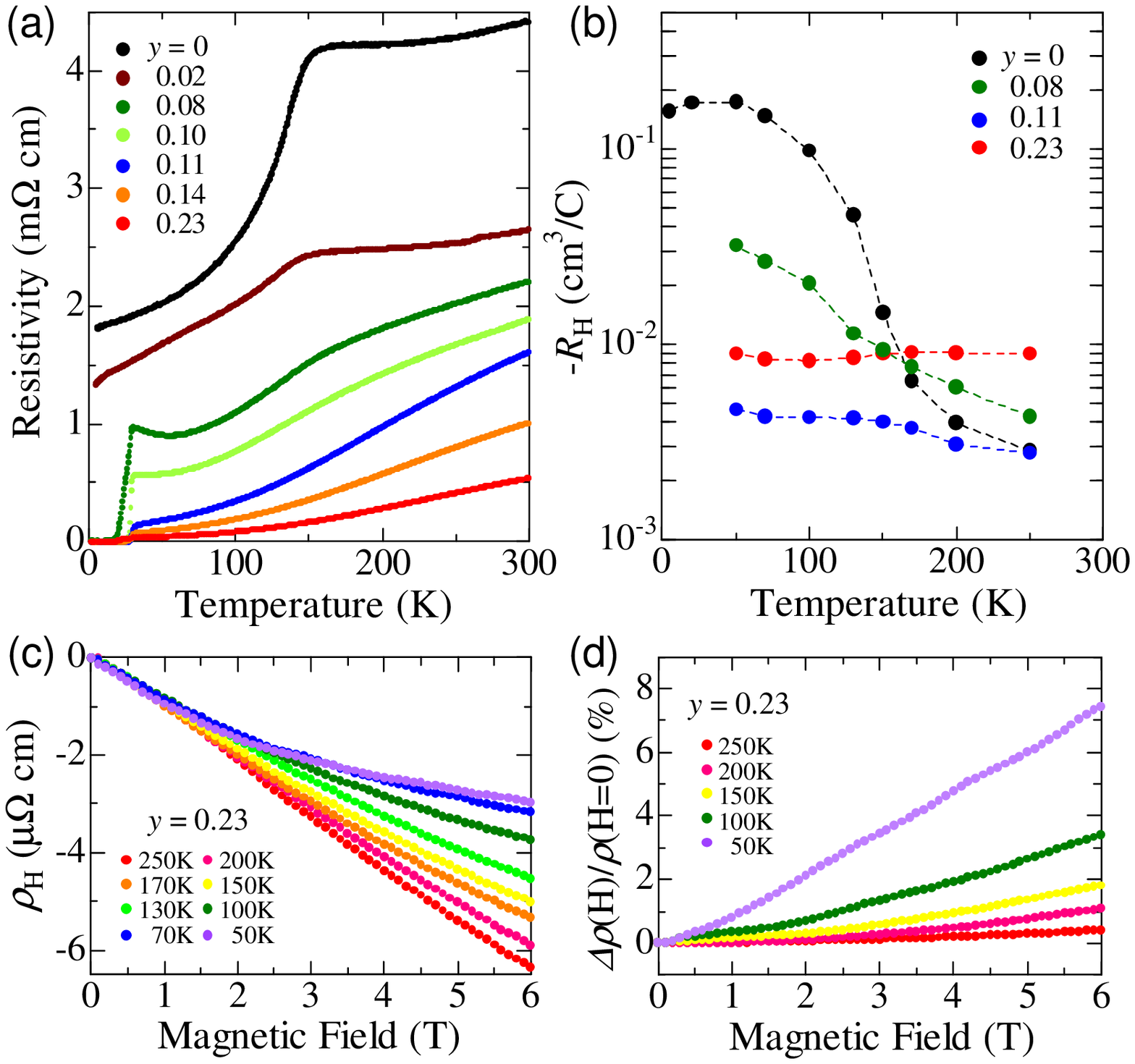}
\caption{\label{fig:1}(color online) (a) Temperature dependence of the resistivity of LaFeAsO$_{1-y}$ for $y$ in the range 0 $\leq y \leq$ 0.23. (b) Temperature variation of the Hall coefficient $R_{\rm H}$ at zero-field limit for four representative samples with $y$ = 0, 0.08, 0.11 and 0.23. Magnetic field dependence of the Hall resistivity $\rho_{\rm H}$($H$) (c) and the transverse magnetoresistance (d) for $y$ = 0.23.}
\end{figure}

\begin{figure}[t!]
\includegraphics[width=0.8\columnwidth,clip]{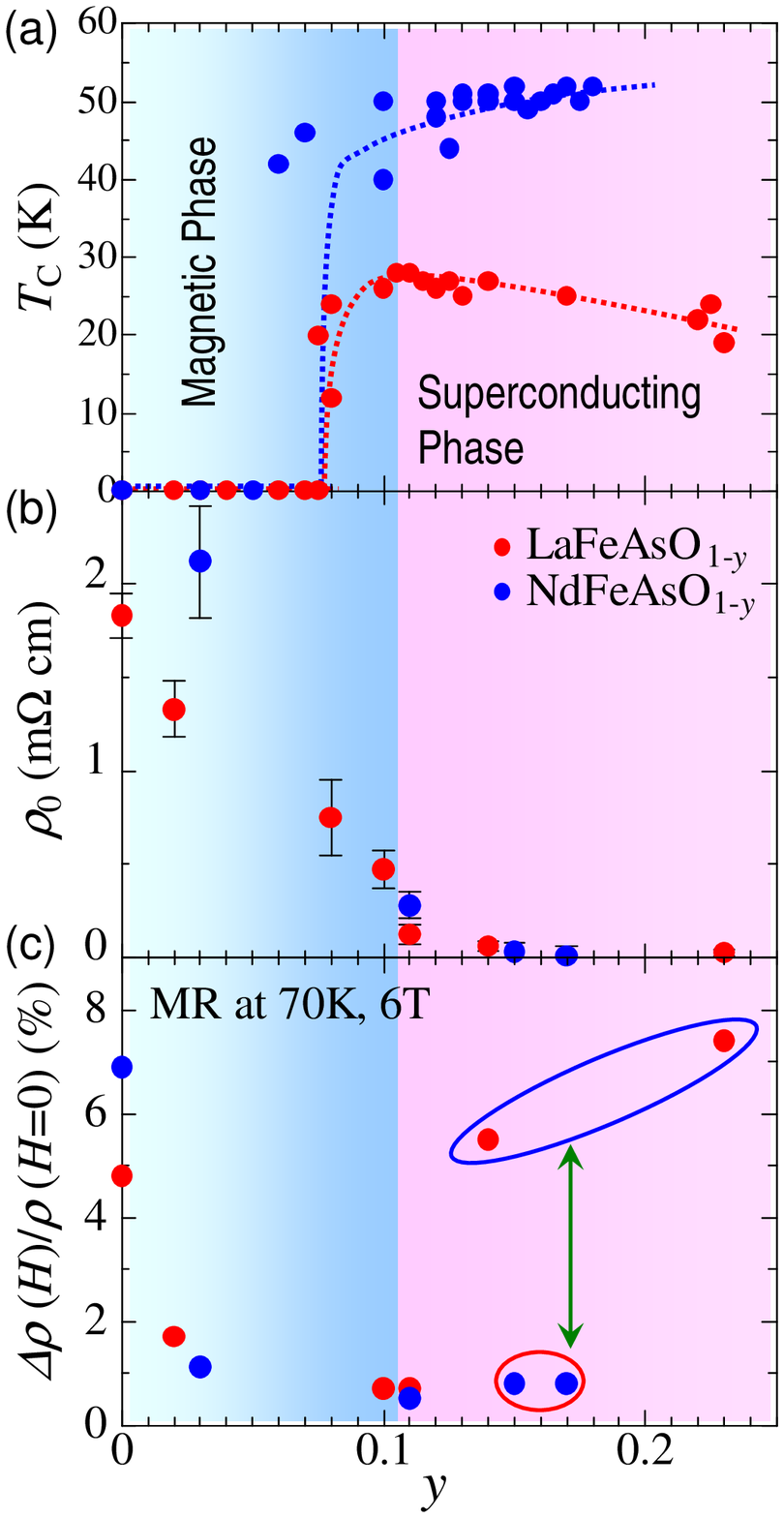}
\caption{\label{fig:2}(color online) (a) Critical temperatures ($T_{\rm c}$) plotted as a function of oxygen deficiency $y$ for LaFeAsO$_{1-y}$ (red circles) and for NdFeAsO$_{1-y}$ (blue circles). $T_{\rm c}$ follows the $y$ dependences reported in previous papers.~\cite{C.H.Lee,Miyazawa} (b) Residual resistivity ($\rho_0$) obtained from fitting (see text) plotted as a function of $y$. (c) The magnitude of magnetoresistance measured at $H$ = 6~T and $T$ = 70~K for both systems.}
\end{figure}

\subsection{LaFeAsO$_{1-y}$}
\indent Figures \ref{fig:1}(a) and (b) show the temperature dependence of resistivity $\rho$($T$) and the Hall coefficient $R_{\rm H}$($T$) for LaFeAsO$_{1-y}$ (0 $\leq y \leq$ 0.23), respectively. $R_{\rm H}$($T$) was determined from the slope of the Hall resistivity $\rho_{\rm H}$ in the low-field limit, i.e. $R_{\rm H}$($T$) = d$\rho_{\rm H}$/d$H$ at $H$ = 0. $\rho$($T$) of the parent compound ($y$ = 0) is only weakly $T$-dependent above 150~K, and it sharply drops upon the antiferromagnetic transition at $T_{\rm N} \sim$ 150~K and shows quadratic $T$-dependence with fairly large residual resistivity at the lowest temperatures. Because the antiferromagnetic order is collinear or stripe-like, the real crystal would consist of domains with different orientations, and scattering at domain boundaries would be a source of residual resistivity. The magnitude of $R_{\rm H}$($T$) rapidly increases below $T_{\rm N}$. The magnitude is larger than that above $T_{\rm N}$ by more than an order of magnitude, and $R_{\rm H}$ was $\sim$ $-$0.2 cm$^3$C$^{-1}$ at low temperatures, corresponding to the nominal electron carrier density of 3 $\times$ 10$^{19}$ cm$^{-3}$. At low temperatures, $R_{\rm H}$ is weakly dependent on $H$, suggesting a contribution of at least two types of carriers, possibly two types of electrons, to $R_{\rm H}$. The transverse magnetoresistance (MR) with the direction of $H$ perpendicular to that of current, $\Delta \rho$($H$)/$\rho$(0) (= [$\rho$($H$) - $\rho$(0)]/$\rho$(0)), is quadratic in $H$ and increases with decreasing temperature. At 5~K it reaches about 8$\%$ at $H$ = 6~T. These results indicate that upon the antiferromagnetic/structural transition, the Fermi surfaces are reconstructed, and the dominant charge carriers continue to be electrons with low carrier density and relatively high mobility.

\indent With increasing $y$, the magnitude of resistivity gradually decreases. Unlike the cuprates, the doping into LaFeAsO does not simply increase the carrier density, as inferred from the nonmonotonic variations of the Hall coefficient with $T$ and $y$ displayed in Fig. \ref{fig:1}(b). A remnant feature of the magnetic/structural transition in $\rho$($T$) for $y$ = 0 seems to remain at least up to $y$ = 0.10, as shown in Fig. \ref{fig:1}(a). This feature is suggestive of the presence of short-range magnetic/structural order which appears to coexist with superconductivity for $y$ = 0.08 and 0.10, and to be a possible source of the large residual resistivity observed for all the compounds with $y$ lower than 0.10. Also, in conjunction with the ``poor'' metallic behavior, the transverse magnetoresistance becomes small over the whole temperature range. Note that $\rho$($T$) exhibits a nearly parallel downshift on going from $y$ = 0.02 to 0.10. This indicates that the dominant effect of doping in this regime is to reduce the residual (elastic scattering) component of resistivity (Fig. \ref{fig:2}(b)) by suppressing the short-range order. We realize that a rapid decrease of the residual resistivity upon doping into the ``parent'' material is a quite common feature observed in almost all FeAs-based superconductors so far investigated.~\cite{L.Fang,Rullier-Albenque,Kasahara,R.H.Liu,Hess} The Hall coefficient in this regime is strongly $T$-dependent. The magnitude of $R_{\rm H}$ for $y$ = 0.08 increases with lowering temperature, just like that for $y$ = 0, possibly due to incomplete Fermi surface reconstruction associated with the short-range order.

\indent When $y$ exceeds 0.10, a remarkable reduction in the residual resistivity takes place (see Fig. \ref{fig:2}(b)), probably as a result of disappearance of the short-range order. Since the amount of oxygen deficiency increases, the reduction of the residual resistivity is evidence that the disorder due to oxygen deficiency in the LaO blocks has a minor effect on carrier scattering. For $y$ $\geq$ 0.11, $i.e.$, in the ``optimal'' and ``overdoped'' regimes, $\rho$($T$) becomes dominated by a $T^2$ term, indicative of electron--electron scattering. Such $T^2$ dependence up to high temperatures is widely observed in correlated electron systems, in which electron--electron interaction dominates the scattering of the carriers.~\cite{Imada} The $T$-dependence of $R_{\rm H}$ becomes weaker. Thus, the charge transport in this regime appears to show normal (good) metallic behavior.

\indent As $y$ further increases from 0.11, the coefficient of the $T^2$ resistivity decreases. In view of the increased magnitude of $R_{\rm H}$ on going from $y$ = 0.11 to 0.23, it is plausible that a decrease of the scattering rate, rather than the increase of carrier density, is responsible for the observed decrease in $\rho$. In fact, the magnitude of MR, a measure of carrier mobility, becomes larger with increasing $y$ (it is 1$\%$ for $y$ = 0.11 and reaches $\sim$ 8$\%$ at $H$ = 6~T and $T$ = 70~K for $y$ = 0.23, as plotted in Fig. \ref{fig:2}(c)). Fig. \ref{fig:1}(c) shows the Hall resistivity for $y$ = 0.23 ($T_{\rm c}$ $\sim$ 19~K). At low fields ($H <$ 1~T), $\rho_{\rm H}$ is linear in $H$ and independent of $T$, $i.e.$, $R_{\rm H}$ is $T$-independent, whereas for $H >$ 1~T, nonlinear behavior is more pronounced as the temperature is lowered. The nonlinear $\rho_{\rm H}$ is clear evidence for the presence of multiple carriers. Judging from the negative sign of $R_{\rm H}$, the nonlinearity of $\rho_{\rm H}$, and the enhanced magnitude of $R_{\rm H}$ with respect to that for $y$ = 0.11, two types of electrons contributed to $\rho_{\rm H}$ (see APPENDIX). One of the two should have higher mobility and lower density, which would have a major contribution to $R_{\rm H}$ and be the origin of the large MR. This is in contrast to the poor-metallic charge transport in the samples with lower $y$'s, or in the ``underdoped'' regime. The charge transport in the ``underdoped'' regime is strongly influenced by disorder (inhomogeneity) produced by remnant magnetic/structural order.

\indent Thus, the charge transport in the LaFeAsO$_{1-y}$ system does not show monotonic evolution with $y$, but shows a distinct behavior in different ``doping'' regimes. The evolution of $\rho$($T$) with $y$ is basically the same as that reported for LaFeAsO$_{1-x}$F$_x$,~\cite{Hess}, although the relation $y$ = 2$x$ expected from a simple valence count does not hold, which seems to be additional evidence showing that the ``doping'' is not a parameter controlling carrier density alone.

\begin{figure}[t!]
\includegraphics[width=\columnwidth,clip]{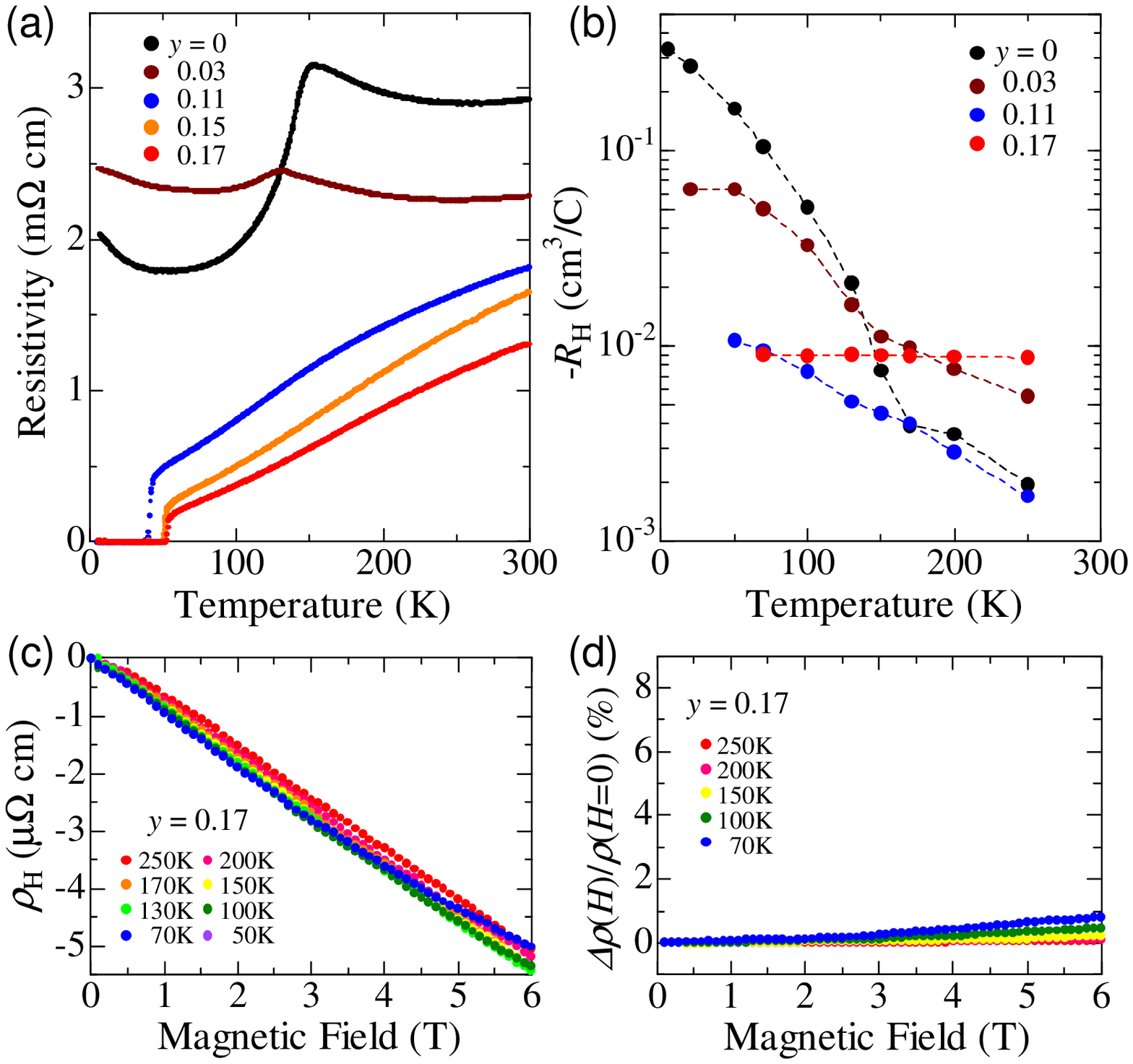}
\caption{\label{fig:3}(color online) (a) Temperature dependence of the resistivity of NdFeAsO$_{1-y}$ for $y$ in the range 0 $\leq y \leq$ 0.17. (b) Temperature variation of $R_{\rm H}$ at zero-field limit for the representative samples with $y$ = 0, 0.03, 0.11 and 0.17. Magnetic field dependence of the Hall resistivity $\rho_{\rm H}$($H$) (c) and the transverse magnetoresistance (d) for $y$ = 0.17.}
\end{figure}

\subsection{NdFeAsO$_{1-y}$}
\indent Figures \ref{fig:3}(a) and (b) display $\rho$($T$) and $R_{\rm H}$($T$) for NdFeAsO$_{1-y}$ (0 $\leq y \leq$ 0.17), respectively. The evolution of these transport properties with $y$ looks qualitatively similar to that for LaFeAsO$_{1-y}$. The residual resistivity sharply diminishes, and the $T$-dependence of $R_{\rm H}$ becomes weaker upon entering the superconducting regime. Although $T_{\rm c}$ continues to rise up to $y$ = 0.17, $R_{\rm H}$ is almost $T$-independent, as in the ``overdoped'' LaFeAsO$_{1-y}$. Similarly, the magnitude of MR rapidly decreased with increasing $y$ (Fig. \ref{fig:2}(c)), showing a minimum at $y$ = 0.11 where $\rho_0$ becomes vanishingly small. However, unlike LaFeAsO$_{1-y}$, MR remains very small for $y$ = 0.15 and 0.17. The most remarkable difference is that the normal-state resistivity in the superconducting regime ($y$ = 0.11, 0.15 and 0.17) is dominated by a $T$-linear term. The $T$-linear resistivity is reminiscent of optimally doped high-$T_{\rm c}$ cuprates, but the $T$-independent $R_{\rm H}$ is in sharp contrast with that for cuprates, in which $R_{\rm H}$ increases with decreasing $T$.~\cite{Takagi}

\indent Figures \ref{fig:3}(c) and (d) show the magnetic-field dependences of the Hall resistivity $\rho_{\rm H}$ and the transverse MR for the $y$ = 0.17 sample. $\rho_{\rm H}$ is linearly dependent on $H$ up to 6~T and does not show an appreciable temperature dependence; that is, $R_{\rm H}$ is independent of both $T$ and $H$. MR is remarkably small even at the lowest temperature (70~K), less than 1$\%$ at $H$ = 6~T. This result suggests that in the $y$ = 0.17 sample, only one type of electron carriers with fairly low density ($\sim$ 7 $\times$ 10$^{20}$ cm$^{-3}$ or $\sim$ 0.05 per Fe) dominates in $R_{\rm H}$ and probably in $\rho$($T$). Other carriers with perhaps comparable or higher density would be ineffective in charge transport due either to the much lower mobility or to gapping in the corresponding Fermi surface. From the small MR, it follows that even the electrons which dominate in $R_{\rm H}$ suffer strong scattering, giving rise to the $T$-linear resistivity.

\subsection{$T_{\rm c}$--$n$ correlation}
\begin{figure}[t!]
\includegraphics[width=0.9\columnwidth,clip]{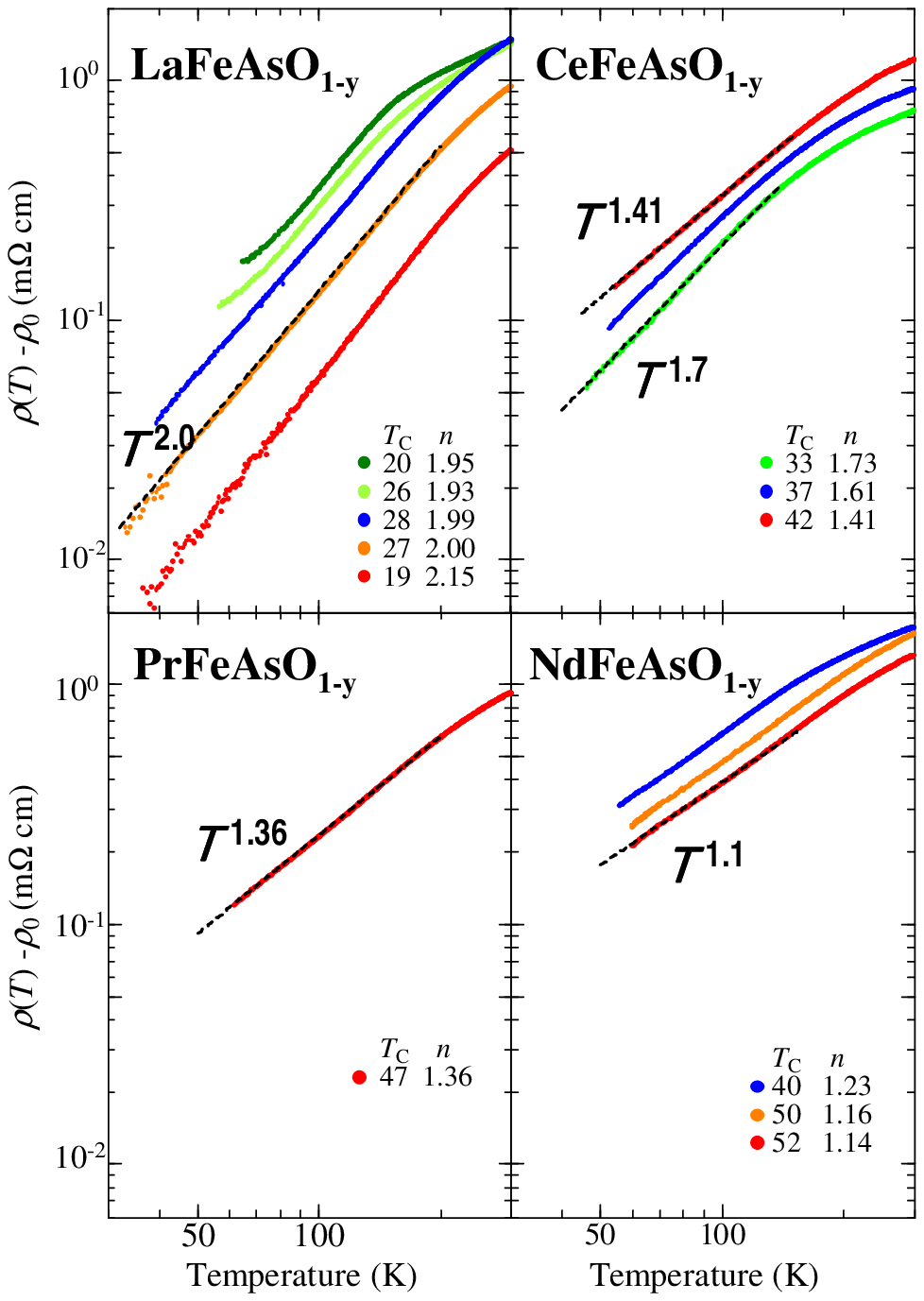}
\caption{\label{fig:4}(color online) log($\rho$($T$) - $\rho_0$) vs. log$T$ plots for LaFeAsO$_{1-y}$, CeFeAsO$_{1-y}$, PrFeAsO$_{1-y}$ and NdFeAsO$_{1-y}$ at various doping levels, where $\rho_0$ is the residual component of resistivity. The and dashed lines are guides for the eye showing the slopes of $\rho$($T$) $\sim$ $T^n$.}
\end{figure}

\begin{figure}[t!]
\includegraphics[width=\columnwidth,clip]{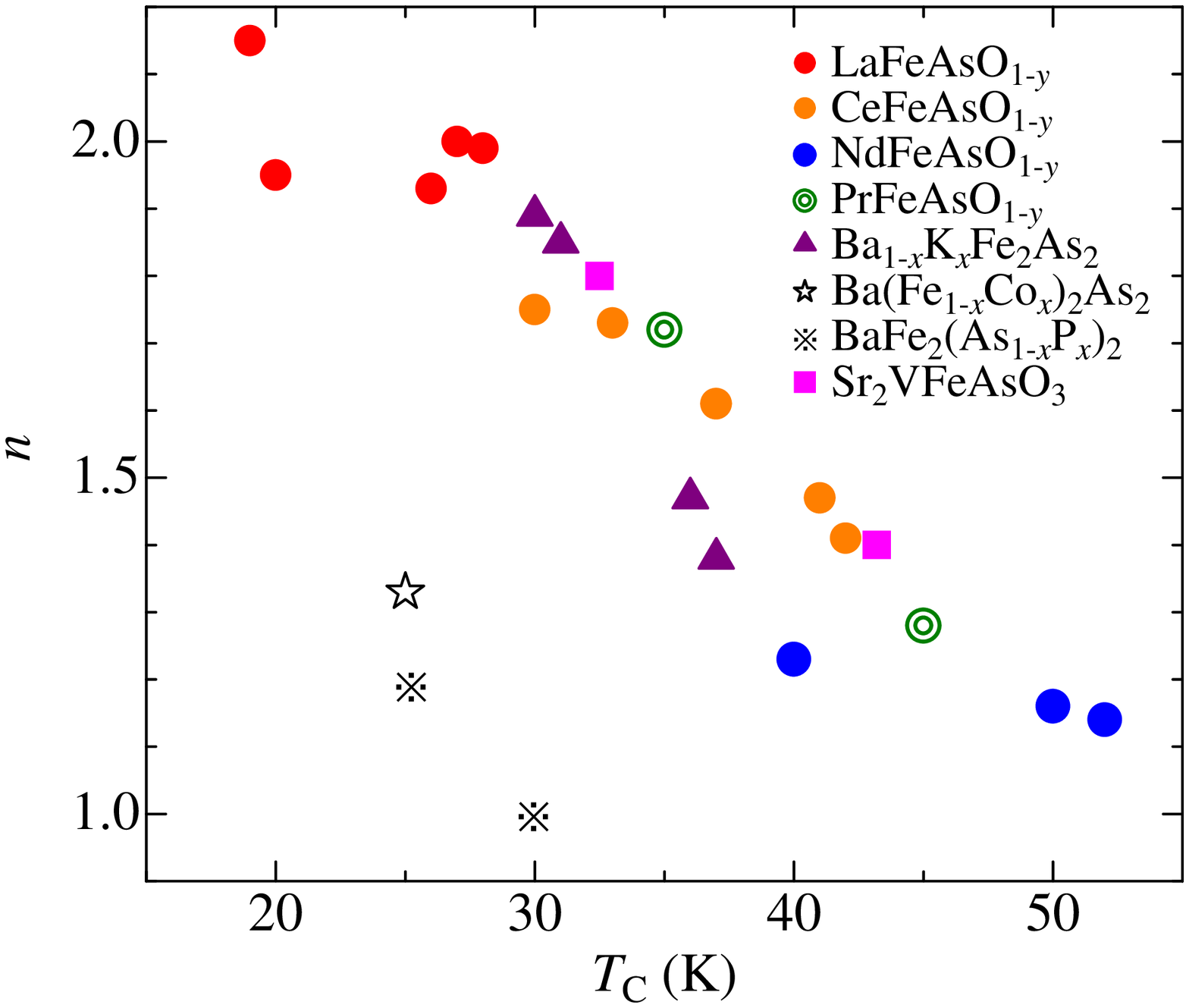}
\caption{\label{fig:5}(color online) Relationship between $T_{\rm c}$ and the exponent $n$ of $\rho$($T$) $\sim$ $T^n$ for various kinds of FeAs-based superconductors. The closed symbols are for LaFeAsO$_{1-y}$ and NdFeAsO$_{1-y}$, and the others are for other systems~\cite{Ishida}, including single crystals of PrFeAsO$_{1-y}$.~\cite{Ishikado} The results for K- and Co- doped BaFe$_2$As$_2$ are in agreement with those reported by other groups. The data for P-doped BaFe$_2$As$_2$ and Sr$_2$VFeAsO$_3$ is from Ref. 11 and 21, respectively.}
\end{figure}

\indent To reveal the correlation between charge transport and $T_{\rm c}$, we investigated the exponent $n$ of $\rho$($T$) $\sim T^n$ for each sample. We fitted the data in the form of $\rho$($T$) = $\rho_0$ + $A$ $T^n$ in the temperature range between $T$ just above $T_{\rm c}$ and $T$ = 150~K (Fig. \ref{fig:4}), where $\rho_0$ is the residual component of resistivity. The exponent $n$ plotted against $T_{\rm c}$ is shown in Fig. \ref{fig:5}. NdFeAsO$_{1-y}$ with higher $T_{\rm c}$ clearly belongs to a class distinct from LaFeAsO$_{1-y}$. On the same figure are also plotted the $n$-values for CeFeAsO$_{1-y}$ and single crystals of PrFeAsO$_{1-y}$ with $T_{\rm c}$ between 30 and 47~K, as well as for K-, Co- and P-substituted BaFe$_2$As$_2$. Except for Co- and P-substituted BaFe$_2$As$_2$, the data points sit in between LaFeAsO$_{1-y}$ and NdFeAsO$_{1-y}$, showing a trend that $n$ decreases from 2 to 1 as $T_{\rm c}$ increases. Note that the compounds showing this $T_{\rm c}$--$n$ correlation have ``dopant'' or disorder sites located outside the FeAs blocks.~\cite{Kamihara,Rotter} $\rho$($T$) of Ba(Fe$_{1-x}$Co$_x$)$_2$As$_2$ and BaFe$_2$(As$_{1-x}$P$_x$)$_2$ is nearly linear in $T$ in their highest $T_{\rm c}$ region, whereas the maximum $T_{\rm c}$ is 25~K in the former and 30~K in the latter. Disorder in the Fe-layer is expected to affect charge transport and $T_{\rm c}$ more seriously. Also, it is known that $T_{\rm c}$ is sensitive to the As-Fe-As bond angle or the pnictogen height from the Fe layer. Therefore, the P-substitution for As sites locally modulate the bond angle / pnictogen height, which would act to reduce $T_{\rm c}$.

\indent The magnitude of MR is always small for superconducting NdFeAsO$_{1-y}$ showing $T$-linear resistivity, whereas MR is much larger in LaFeAsO$_{1-y}$ with resistivity dominated by a $T^2$ term. Thus, one may conclude that strong ``inelastic'' carrier scattering is the origin of the $T$-linear resistivity in the normal state and a possible ingredient for achieving $T_{\rm c}$ of 40--50~K in Fe-based superconductors. Electron--phonon interaction is probably not the origin of the $T$-linear resistivity in the Fe-based superconductors, as it is unlikely that the electron--phonon scattering gives rise to $T$-linear resistivity in some cases and to $T^2$ resistivity in others in the same class of compounds.~\cite{Boeri}

\indent This correlation between the exponent $n$ and $T_{\rm c}$ is not restricted to $Ln$FeAsO$_{1-y}$ investigated in the present work.~\cite{Medvedev,Margadonna} In the case of LaFePO, $T_{\rm c}$ is as low as $\sim$7~K, and $\rho$($T$) obeys a $T^2$-dependence over a wide $T$ range.~\cite{Sugawara,Carrington} A crossover from $T^2$ to $T$-linear dependence of resistivity associated with an increase of $T_{\rm c}$ is seen in one system, La$_{1-x}$Y$_x$FeAsO$_{1-y}$, in which $T_{\rm c}$ goes up to 43~K with $x$,~\cite{Shirage} and also in NdFeAsO$_{1-y}$ under pressure where $n$ increases from about 1 to 2 as $T_{\rm c}$ decreases with pressure.~\cite{Takeshita}

\subsection{Comparison with theories}
\indent In the band structure of nondoped LaFeAsO,~\cite{Kuroki2} the top of a hole band (called a $\gamma$ band) centered at $\rm{\Gamma}$ is located just above the Fermi level. It moves down with electron doping and finally is located below the Fermi surface. In the case of NdFeAsO$_{1-y}$ the top of the $\gamma$ band is higher than that for LaFeAsO$_{1-y}$ being always above the Fermi level regardless of electron doping level. It is argued that a Fermi surface derived from the $\gamma$ band facilitates nesting between disconnected portions of electron Fermi surfaces, which would enhance ``interband'' scattering of electrons between the two nesting Fermi surfaces and, in certain circumstances ($e.g.$, when coupled with spin fluctuations), would be a source of pairing interaction.~\cite{Mazin,Kuroki1} In addition, the location of the $\gamma$ band is found to be sensitive to local crystal structure, specifically, the As height from the Fe plane or the bond angle $\alpha$ of the FeAs$_4$ octahedron.~\cite{Kuroki2} Then, the $\gamma$-Fermi surface seems to bridge between the known $T_{\rm c}$-$\alpha$ correlation and the present $T_{\rm c}$-$n$ correlation we observed. Unfortunately, there is no experimental evidence for the presence of the $\gamma$-Fermi surface in NdFeAsO$_{1-y}$ and its absence in LaFeAsO$_{1-y}$.

\indent An alternative scenario is that NdFeAsO$_{1-y}$ showing $T$-linear resistivity is located near some quantum critical point (QCP), and electrons are subject to strong scattering from quantum critical fluctuations, as is discussed in the heavy fermion systems for a source of $T$-linear resistivity. A possible QCP in Fe-based systems is the point at which the magnetic or structural order, either long- or short-range order, disappears on the doping axis. In the present case, it would be $y$ $\sim$ 0.1, where the residual resistivity sharply drops and high-$T_{\rm c}$ superconductivity sets in. Then, a question would arise: what makes LaFeAsO$_{1-y}$ distinct from NdFeAsO$_{1-y}$ in that the $T^2$ term dominates in resistivity even in the vicinity of $y$ $\sim$ 0.10?

\section{\label{sec:level4}Conclusion}
\indent The present results strongly suggest that the strong carrier scattering which gives rise to $T$-linear resistivity is intimately correlated with high-$T_{\rm c}$ in the Fe-based superconductors. In conventional BCS superconductors, the electron--phonon scattering rate, which determines the normal state resistivity, is linked to the strength of the pairing interaction in the superconducting state. A correlation between $T$-linear resistivity and high $T_{\rm c}$ is also discussed in the context of quantum criticality in organic, heavy fermion, and cuprate superconductors, as a common feature of the normal state of unconventional superconductors.~\cite{Doiron-Leyraud} However, in the case of cuprates, the carrier scattering rate is not necessarily related with $T_{\rm c}$. $T_{\rm c}$ is controlled by the superfluid density, and the transport scattering rate is not strongly dependent on doping and material in the underdoped and optimally doped regimes. In view of the charge transport demonstrated in the present work, the Fe-based superconductors apparently belong to a class different from the cuprates, and seem to have an aspect in common with other classes of superconductors in that the carrier scattering has some relevance to $T_{\rm c}$.




\section*{\label{sec:level5}ACKNOWLEDGMENTS}
\indent We would like to acknowledge K. Kuroki, H. Aoki, S. Ishibashi, and T. Miyake for helpful discussions. This work was supported by the Global Centers of Excellence Program  and A3 Foresight Program from the Japan Society for the Promotion of Science and Transformative Research-Project on Iron Pnictides from the Japan Science and Technology Agency.\\

\section*{\label{sec:level6}APPENDIX}

\begin{figure}[t!]
\includegraphics[width=0.9\columnwidth,clip]{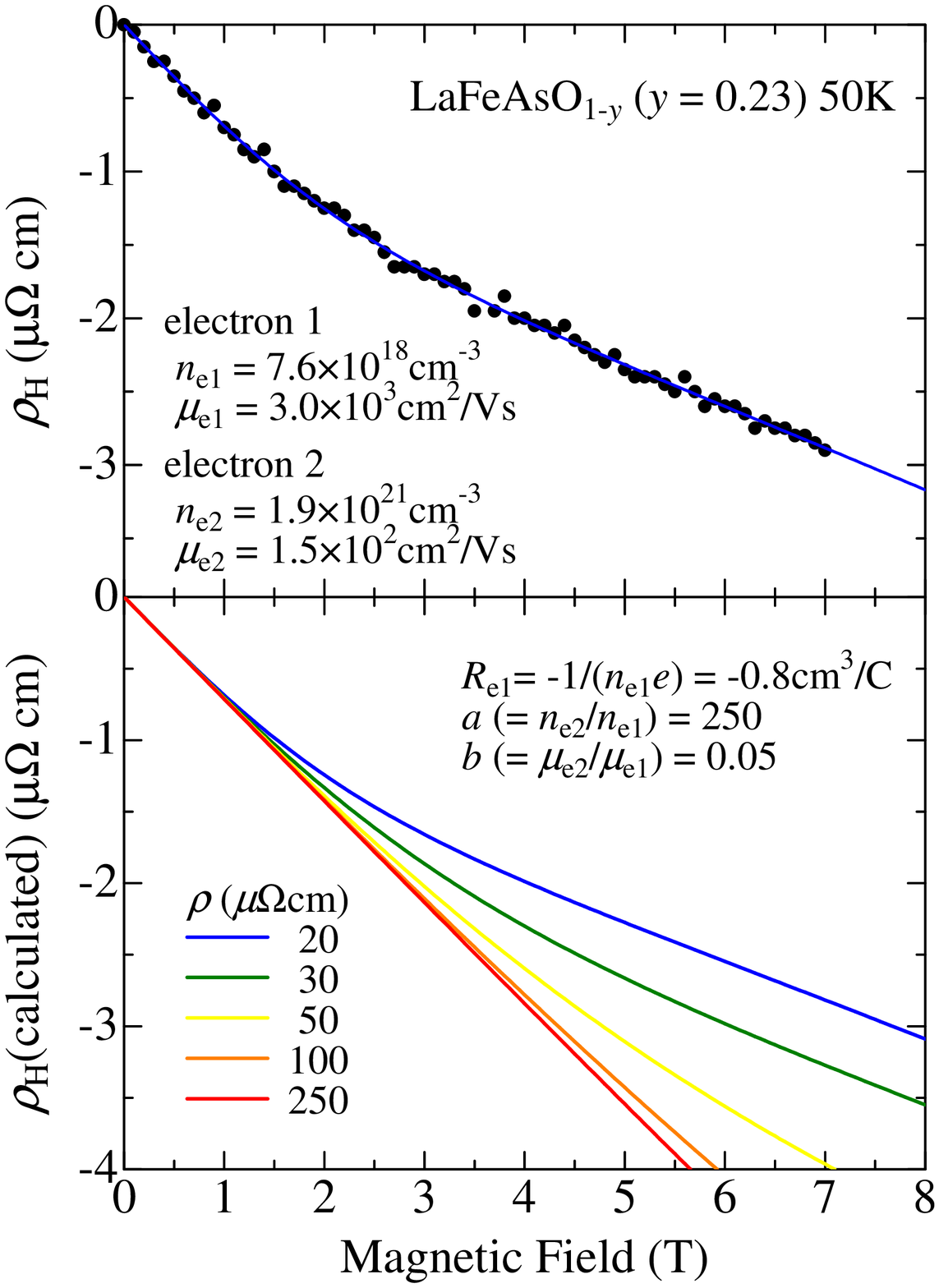}
\caption{\label{fig:A}(color online) (a) The magnetic field dependence of Hall resistivity $\rho_{\rm H}$ at 50~K for LaFeAsO$_{1-y}$ ($y$ = 0.23). The blue line was obtained by fitting assuming a two-carrier model. (b) The calculated results for some temperatures.}
\end{figure}

\indent By fitting $\rho_{\rm H}$($H$) assuming a two-carrier model, we obtained a result that two types of electron carriers reproduce the data  in Fig. \ref{fig:1}(c), as follows. Namely, $n_{e1}$ $\sim$ 7.6 $\times$ 10$^{18}$~cm$^{-3}$, $\mu_{e1}$ $\sim$ 3.0 $\times$ 10$^3$~cm$^2$V$^{-1}$s$^{-1}$,  $n_{e2}$ $\sim$ 1.9 $\times$ 10$^{21}$~cm$^{-3}$, and $\mu_{e2}$ $\sim$ 1.5 $\times$ 10$^2$~cm$^2$V$^{-1}$s$^{-1}$ at $T$ = 50~K, where $n_{ei}$, $\mu_{ei}$, and $i$ represent carrier number, mobility, and type of carriers ($i$ = 1 and 2), respectively. The temperature dependence of $\rho_{\rm H}$($H$) is also reproducible, assuming the same temperature dependence of both $\mu_{ei}$'s as that of the inverse of resistivity. Note that the existence of the minor carrier ($i$ = 1) with high mobility enhances the absolute value of $R_{\rm H}$ at the low-field limit, which is larger than that simply expected for the major carrier ($i$ = 2). Although the obtained parameter set is somewhat unrealistic and the origin of the minor carrier is not clear from a band calculation, the nonlinearity of $\rho_{\rm H}$($H$) may be related to the configuration of Fermi surfaces.

\end{document}